\documentclass{jfm}
\usepackage{psfig}
\usepackage{amssymb}
\usepackage{amsbsy}

\def\eg{{e.g.\ }}

\def\etal{\mbox{\it et al.\ }}

\title[Passive scalar anisotropy and LES]
{Passive scalar anisotropy in a heated turbulent wake: 
new observations and implications for LES}

\author[Hyung Suk Kang and Charles Meneveau]%
{H\ls Y\ls U\ls N\ls G\ns S\ls U\ls K\ns K\ls A\ls N\ls G$^1$
 \and C\ls H\ls A\ls R\ls L\ls E\ls S\ns M\ls E\ls N\ls E\ls V\ls E\ls A\ls
U$^1$}

\affiliation{$^1$Department of Mechanical Engineering, Johns Hopkins
University,
3400 North Charles Street, Baltimore, MD 21218, USA\\[\affilskip]}

\pubyear{2001}
\volume{777}
\pagerange{1--126}
\date{?? and in revised form ??}
\begin{document}

\maketitle

\begin{abstract}

The effects of passive scalar anisotropy on
subgrid-scale (SGS) physics and modeling for Large-Eddy Simulations are studied
experimentally. Measurements are performed across a moderate Reynolds number wake flow
generated by a heated cylinder, using an array of four X-wire and four cold-wire probes. 
By varying the separation distance among probes
in the array, we obtain filtered and subgrid quantities at three different
filter sizes.  We compute several terms that comprise the subgrid dissipation
tensor of kinetic energy and scalar-variance and test for isotropic
behavior, as a function of filter scale. We find that whereas the kinetic 
energy dissipation tensor tends towards isotropy at small scales, the
SGS scalar-variance dissipation remains anisotropic independent of
filter scale. The eddy-diffusion model predicts isotropic
behavior, whereas the nonlinear (or tensor eddy diffusivity) model
reproduces the correct trends, but overestimates the level of scalar dissipation
anisotropy. These results provide some support for so-called mixed models but raise
new questions about the causes of the observed anisotropy.
\end{abstract}

\section{Introduction}

The statistics and general structure of passive scalars in turbulent flows
differ significantly from those of the turbulent velocity field. In
particular, conclusive experimental
(e.g. Stewart 1969, 
Sreenivasan, Antonia \& Britz 1979, Mestayer 1982, and Mydlarski \& Warhaft 1998a)
and numerical (\eg Holzer \& Siggia 1994) evidence
shows that
structure functions and the derivative skewness of the scalar
field do not follow predictions from
isotropy at inertial and dissipative scales, in the presence of a mean scalar
gradient. In
particular, the deviations are thought to be related to ``ramp and
cliff structures''  and to imply a direct effect of large-scale structures on
small-scale structures. The
data relevant to this question has been reviewed by Sreenivasan (1991) and
more recently by
Warhaft (2000).  Moreover, scalar spectra display a distinctly less
universal structure than
velocity spectra. This is manifested both in terms of spectral exponents,
as well as in terms of the dimensionless spectral coefficient $c_{\theta}$. 
The latter varies from
values near 0.4 in many
experiments in the atmospheric surface layer and grid turbulence (see
Sreenivasan 1991, Sreenivasan 1996 and Warhaft 2000) to
$c_{\theta} \sim 1.8$ for other atmospheric measurements 
(Antonia, Ould-Rouis, Anselmet \& Zhu 1997, and Antonia, Xu \& Zhou 1999).
From a fundamental point of view, these
observations challenge the Kolmogorov cascade phenomenology for the
transfer of scalar-variance
from large to small scales. In the classical phenomenology, the
multiplicity of separate ``eddy breakdown'' events is assumed to gradually
uncouple the small from the large scales, allowing the former to tend to a
universal and isotropic structure more or less independent of the large
scales. On the contrary, the scalar-field measurements suggest a direct
linkage among largest and smallest scales. 

The coupling among scales is an important ingredient in Large-Eddy Simulations (LES).
In LES, the turbulent fields (velocity and scalar) are
decomposed into large and small (subgrid-scale, SGS) scale contributions by means of a
spatial low-pass filter of characteristic width  $\Delta$. The resulting equations 
which can be numerically discretized with mesh-spacing of the order of $\Delta$ require
closure of the unresolved momentum fluxes (SGS stress tensor, $\tau_{ij} \equiv
\widetilde{u_i u_j} - \widetilde{u}_i\widetilde{u}_j$, where a {\it tilde} represents
filtering at scale $\Delta$) and scalar fluxes (e.g. the SGS heat flux, $q_{j} \equiv
\widetilde{\theta u_j} - \widetilde{\theta} \widetilde{u}_j$
where $\theta$ is the passive scalar field). The promise of LES is often predicated 
upon small-scale universality and isotropy, and the absense of a strong coupling
across disparate length scales. Hence, the observed anisotropy of the scalar field
seems to pose a challenge to the very foundation of LES. While these 
deviations from classical phenomenology are now quite well established, little is
known about their impact on the closure problem for LES. The present work
quantifies the implications of small-scale scalar anisotropy on quantities that
describe subgrid-scale physics and directly affect modeling for LES. 

As reviewed in Meneveau \& Katz
(2000), the most important statistical property of the fluxes $\tau_{ij}$ and $q_j$
is  how they affect the mean kinetic energy and scalar-variance budgets of the
resolved fields.  Specifically, their dominant effect is through the kinetic energy
and scalar-variance dissipations that arise from interactions between subgrid and
resolved scales.  Therefore, in the present study, we mainly focus on the
so-called SGS kinetic energy dissipation $-\langle\tau_{ij}\widetilde{S}_{ij}\rangle$
(Piomelli \etal, 1991) and scalar-variance dissipation 
$-\langle q_{j} \widetilde{G}_j\rangle$
(Port\'e-Agel, Meneveau \& Parlange 1998).
Here  $\widetilde{S}_{ij} \equiv
\frac{1}{2} \big( \partial\widetilde{u}_i / \partial x_j +
\partial\widetilde{u}_j / \partial x_i \big) $
and $\widetilde{G}_j \equiv
\partial \widetilde{\theta} / \partial x_j $
are the resolved strain-rate tensor and scalar gradient, respectively.
The SGS dissipation rates represent the flux (cascade) of kinetic energy or
scalar-variance from resolved towards subgrid scales (when positive). 
When $\Delta$ pertains to the inertial range, and when the flow is 
in equilibrium, one expects the mean SGS dissipation to equal 
the molecular dissipation rate. 

Deviations from isotropy in the context of SGS dissipation can be 
probed by measuring the isotropy level of the tensors
$-\langle\tau_{ij}\widetilde{S}_{mn}\rangle$ and
$-\langle q_{i}\widetilde{G}_j\rangle$, as a function of scale.  The
main question to be addressed in this work is whether the approach to isotropy
(if it exists) is the same for kinetic energy and scalar-variance
dissipation tensors. Another goal is to test the ability of 
two popular model classes (eddy diffusivity and nonlinear models, see
Meneveau \& Katz 2000 for a review) to reproduce the observations. 
In order to span a sizeable range of inertial-range filter scales, a
sufficiently high Reynolds number must be considered. Hence, this study is
based on experimental data (as opposed to DNS which is limited to small
Reynolds numbers). The study is performed in a canonical shear flow, 
the heated cylinder wake. 

\section{Experiment apparatus and flow characteristics}
Experiments were performed in the  return type Corrsin Wind Tunnel 
(Comte-Bellot \& Corrsin 1966). A heated smooth cylinder of diameter $D$ =
4.83 cm was located horizontally at the centerline
of the test section. The measurement location in the streamwise direction $(x_1)$ was
fixed at $x_1/D=25$. To obtain the filtered and SGS quantities, an array of four
custom-made miniature probes was used. Each probe was composed of one X-type
hot-wire and one I-type cold-wire for the velocities in the $x_1-x_2$ plane and
the temperatures, respectively. Here $x_2$ is the ``cross-wake'' direction,
i.e. perpendicular to $x_1$ and to the cylinder axis.

The separation distance $h$ between the probes in the cross-wake direction $x_2$
could be adjusted manually between 5 and 20 mm. Three configurations
($\Delta=2h$, with $h=5$, 10 and 20 mm) were used in the present study. A 2.5 $\mu$m
platinum-coated tungsten wire which had been copper-plated was soldered on to the
X-wire prong ends and etched, yielding an active length-to-diameter ratio of
about 200.  The wire spacing between the hot-wires
was 0.5 mm. A 0.625 $\mu$m silver-coated pure platinum wire for a cold-wire sensor was
soldered on the I-wire prong ends and subsequently etched. To minimize the low
frequency amplitude attenuation, the active length-to-diameter ratio was about 1000
as suggested by \cite{Bruun95}. The separation distance between the cold-wire and
its nearest hot-wire was 0.9 mm so that the thermal effect from the hot-wire
on the cold-wire was negligible. The signals were low-pass filtered at a frequency of
20 kHz and sampled at $f_s = 40$ kHz. Sampling time was 60
seconds, so the total number of data points per channel for each measurement location
was $2.4\times10^6$. The array was traversed across the wake, and data were
recorded at 17 discrete cross-wake locations from the centerline to the wake
edge at increments of 14.4 mm.

At the measurement location of $x_1/D=25$,
the mean centerline velocity $(U_{CL})$ was $13.6$ m/s,
the defect velocity $(U_{d}=U_{\infty}-U_{CL})$ was $4.4$ m/s,
the defect temperature $(\theta_d=\theta_{CL}-\theta_{\infty})$ was $0.61
~^{\circ}$C, and the half-width of the wake was $\ell = 0.08$ m.
To get the spatial quantities along the streamwise direction from the
temporal data, Taylor's hypothesis was invoked.
The turbulence intensity of the streamwise velocity at the centerline was
about $13.3\%$. 
The molecular kinetic energy dissipation at the centerline 
$(\epsilon_{CL})$ was $87$ m$^{2}$s$^{-3}$, and the
molecular scalar-variance dissipation at the centerline
$(\epsilon_{\theta CL})$ was $0.65~^{\circ}$C$^2$s$^{-1}$. The latter two
variables were obtained from (corrected) third-order structure functions
as in Cerutti \etal (2000) and \cite{Lindborg99}. It
follows that the Kolmogorov length scale
$(\eta=(\nu^3/\epsilon)^{1/4})$ was
$0.08$ mm, the Taylor micro-scale $(\lambda)$ was
$2.9$ mm, and the Reynolds number based on Taylor micro-scale
$(Re_{\lambda})$ was $350$.
The longitudinal integral scale obtained
by integrating up to the first
zero crossing of the $u_1$ correlation function was $L_{11} = 0.091$ m.
Profiles of mean velocity, rms velocities, Reynolds shear stress, rms
temperature and heat flux distributions across the wake
agreed quite well with results in the literature
(e.g., Matsumura \& Antonia 1993 and Kiya \& Matsumura 1988).

In the present study, to separate between large and small scales,
the box filter is
applied to  the streamwise and cross-wake directions, and
the trapezoidal rule is used for the spatial integrations.
The filtering process consists in a discrete approximation to a two-dimensional box filter. 
In the $x_2$ direction, a four point discretization is used for evaluating 
the SGS fluxes while a
three-point approximation is used for the filtered derivatives.   
Filtered velocity and scalar gradients in the $x_2$ direction are evaluated 
using first-order finite differences over a distance $h$. In the streamwise direction, 
the box filter is approximated using $\Delta f_s/\langle u_1\rangle$ sampling points 
and the $x_1$ derivatives are evaluated using finite differences over a distance $h$. 
The filtering and error analysis is documented 
in Cerutti \& Meneveau (2000) and Cerutti {\it et al.} (2000).

\begin{figure}
\centerline{\hbox{
\psfig{figure=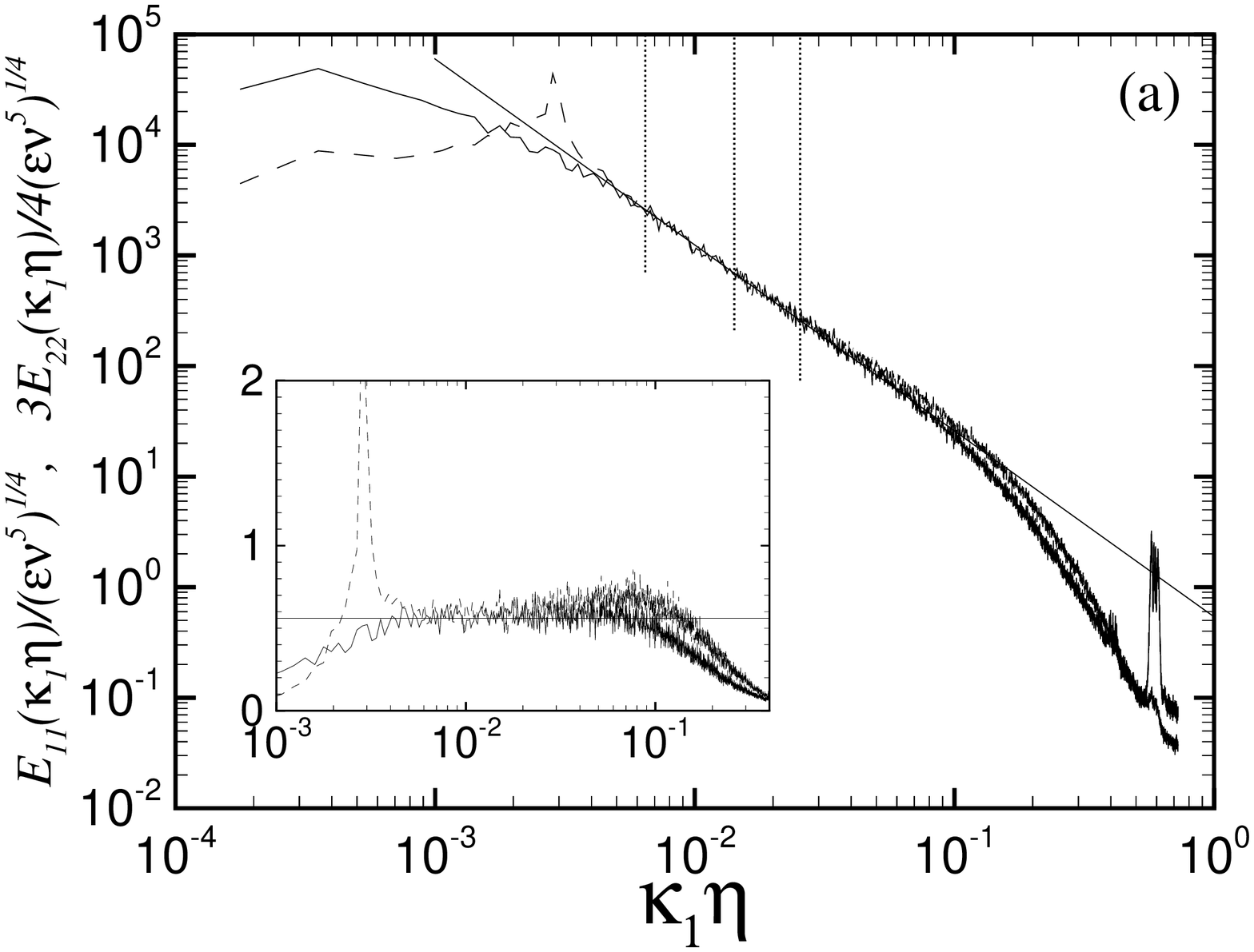,width=7cm,angle=0}
\psfig{figure=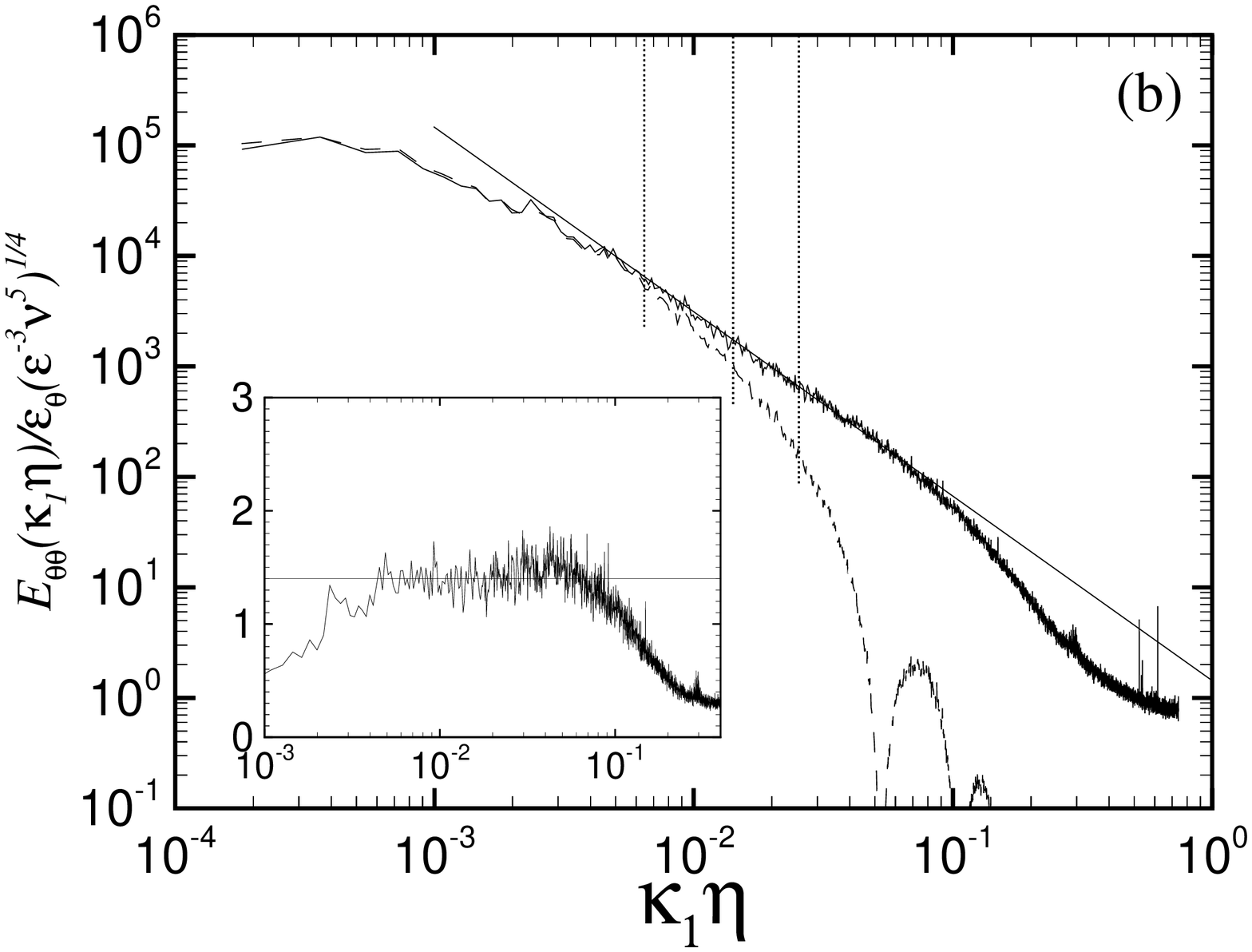,width=7cm,angle=0}}}
\caption{Velocity and temperature spectra at the centerline, in Kolmogorov
units: (a) the longitudinal spectrum of the $u_1$ component (solid line)
and the longitudinal spectrum of the $u_2$ component multiplied by 3/4 (dashed line),
as a function of the longitudinal wavenumber, $\kappa_1$.
The spectrum of $u_2$ component is below that of $u_1$ at low wave
number ($\kappa_1 \eta<0.002$) with a Strouhal peak ($\kappa_1 \eta \sim
0.003$), and higher at high wave number ($0.1<\kappa_1 \eta<0.4$);
(b) the longitudinal spectrum of the temperature (solid line) and 
of the filtered temperature, filtered at $\Delta / \eta=125$ (dashed line). 
The inserts in (a) and (b) show the compensated spectra,
$E_{11}(\kappa_1) ~\epsilon^{-2/3}\kappa_1^{5/3}$, 
$\frac{3}{4}E_{22}(\kappa_1)~ \epsilon^{-2/3}\kappa_1^{5/3}$
and $E_{\theta\theta}(\kappa_1)~ \epsilon_{\theta}^{-1} \epsilon^{1/3}\kappa_1^{5/3}$.
The three vertical dotted lines in (a) and (b) are the wavenumbers corresponding 
to the filter sizes of $\Delta / \eta=$125 (10 mm), 250 (20 mm), 500 (40 mm). 
The straight solid lines are the universal spectra (see text).}
\label{<<Fig1>>}
\end{figure}

Figure 1(a) shows a comparison between the longitudinal spectrum
$E_{11}(\kappa_1)$ of the $u_1$ component  and the longitudinal spectrum
$E_{22}(\kappa_1)$ of the $u_2$ component  multiplied by 3/4 at the
centerline. Here, $\kappa_1$ is the longitudinal wave number.
The three vertical dashed lines correspond to the filter sizes of
${\Delta / \eta =}$125 (10 mm), 250 (20 mm), 500 (40 mm).
All the filter sizes are in the inertial range.
Since the noise peak in the longitudinal spectrum of $u_1$ 
is in the far dissipation region,
quite removed from any of the filter frequencies and scales of interest in this
study, no effort is made to remove the noise by additional filtering (various
attempts such as notch filtering showed no effect on the results). 
It can be clearly observed that $E_{11}(\kappa_1) \approx {3\over 4}
E_{22}(\kappa_1)$ as required by isotropy in the inertial range, over about
one decade of wave numbers. The peak in $E_{22}(\kappa_1)$ at
$\kappa_1 \eta \sim 0.003$ is due to the periodic von K\'arm\'an vortex street
behind the cylinder. The frequency of the vortices is 78.1 Hz,
and this gives Strouhal number $S_t~(=fD/U_\infty) = 0.211$.
The Kolmogorov constant $c_K$  in
$E_{11}(\kappa_1)=\frac{18}{55}c_K \epsilon^{2/3}\kappa_1^{-5/3}$ is 
obtained as $c_K=\frac{55}{18}(0.56) = 1.71$ in the insert. This value is
quite close to the standard value of 1.6 (see review by Sreenivasan 1995)
and a similar value of 1.7 was observed by O'Neil \& Meneveau (1997).
Figure 1(b) shows the longitudinal spectrum of the temperature $E_{\theta
\theta}(\kappa_1)$. 
As seen in the premultiplied spectrum shown in the insert,
there is a fairly clear inertial range with a $-5/3$ slope.
This slope is steeper than those found in the round jet by Tong \& Warhaft (1995) 
and Miller \& Dimotakis (1996), and in several other shear flows reviewed 
in Sreenivasan (1996). 
It is closer to results quoted in Antonia \& Pearson (1997), 
who report a scaling exponent of
0.65-0.66 for the $2^{nd}$-order temperature structure function in the heated cylinder wake 
at $Re_{\lambda} = 230$, or to the spectra for grid turbulence of 
Mydlarski \& Warhaft (1998b).

The coefficient $c_{\theta}$ in 
$E_{\theta\theta}(\kappa_1)= c_{\theta} \epsilon_{\theta} \epsilon^{-1/3}\kappa_1^{-5/3}$
deduced from this spectrum is about 1.4. It is significantly higher than results
quoted in Sreenivasan (1996) for high Reynolds numbers, but is within the range of results 
from various shear flow measurements  reported in Antonia \etal
1999. As discussed in Sreenivasan (1996), the dissipation measurements are based on
assuming scalar isotropy and are thus subject to considerable uncertainty. On the other hand,
at the centerline we find good isotropy (see section 3) which would seem to support
our current estimates of the dissipation and $c_{\theta}$. While the universality of
$c_{\theta}$ and spectral exponent is not the main subject of this paper, the scatter of
results certainly supports the view that the scalar spectrum for shear flows at
moderate Reynolds numbers (e.g. $Re_{\lambda}<1000$) depends upon details of the generation of the
flow (Sreenivasan, 1996). Specifically, it seems that the dependence of the spectral exponent and
prefactor on $Re_{\lambda}$ is not universal.

The longitudinal spectrum of the filtered temperature 
is shown as the dashed curve in figure 1(b). 
The lobes at scales below the filter size are due to the
streamwise box filter used in the present study. 
Spectra of filtered velocity (not shown) have similar shape.

\section{Isotropy of real SGS dissipation and model predictions}
\begin{figure}
\centerline{\hbox{
\psfig{figure=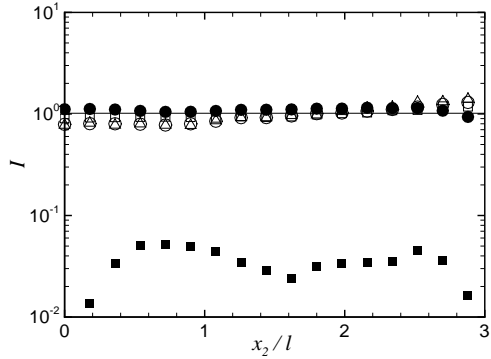,width=6.7cm,angle=0}}}
\caption{Isotropy ratios of the filtered gradient fields and of the
mean SGS stresses across the heated wake at $x_1/D=25$ 
with the filter size of $\Delta/\eta=125$.
The symbols represent:
{\Large{$\circ$}}, 
$\langle\widetilde{S}_{22}\widetilde{S}_{22}\rangle / 
\langle\widetilde{S}_{11}\widetilde{S}_{11}\rangle$;
$\square$, 
$\langle\widetilde{S}_{12}\widetilde{S}_{12}\rangle 
/ \langle\frac{3}{4}\widetilde{S}_{11}\widetilde{S}_{11}\rangle$;
$\vartriangle$,
$\langle\widetilde{G}_{2}\widetilde{G}_{2}\rangle / 
\langle\widetilde{G}_{1}\widetilde{G}_{1}\rangle$;
{\Large{$\bullet$}}, 
$\langle\tau_{22}\rangle / \langle\tau_{11}\rangle$;
$\blacksquare$, 
$-\langle\tau_{12}\rangle / \langle\tau_{11}\rangle$.
}
\label{<<Fig2>>}
\end{figure}

In discussing isotropy of various tensors, we distinguish second-rank and
fourth-rank tensors. Examples of second-rank tensors are the mean SGS stress
$\langle\tau_{ij}\rangle$, the filtered scalar-gradient
$\langle\widetilde{G}_{i}\widetilde{G}_{j}\rangle$, 
and the SGS scalar-variance dissipation
$-\langle q_{i}\widetilde{G}_{j}\rangle$. Their isotropic form can be written as
$\langle\tau_{ij}\rangle = \langle\tau_{11}\rangle \delta_{ij}$,
$\langle\widetilde{G}_{i}\widetilde{G}_{j}\rangle = 
\langle\widetilde{G}_{1}\widetilde{G}_{1}\rangle \delta_{ij}$, and
$\langle q_{i}\widetilde{G}_{j}\rangle = 
\langle q_{1}\widetilde{G}_{1}\rangle \delta_{ij}$ respectively.
On the other hand, the strain-rate-product tensor
$\langle\widetilde{S}_{ij}\widetilde{S}_{pq}\rangle$ and  
SGS dissipation tensor $\langle\tau_{ij}\widetilde{S}_{pq}\rangle$
are fourth-rank tensors. Their isotropic form can be written as
\begin{equation}
\left<\tau_{ij}\widetilde{S}_{pq}\right> =
-\frac{1}{2} \left<\tau_{11}\widetilde{S}_{11} \right>
\left[ \delta_{ij}\delta_{pq}-
\frac{3}{2} \left(\delta_{ip}\delta_{jq}+\delta_{iq}\delta_{jp}\right)
\right],
\end{equation}
and a similar expression holds for 
$\langle\widetilde{S}_{ij}\widetilde{S}_{pq}\rangle$
by replacing $\tau_{ij}$ with $\widetilde{S}_{ij}$. To derive the
above expression, we have used the tensor symmetry (in $i-j$ or $p-q$) and
the divergence-free condition ($\widetilde{S}_{kk} = 0$).
It follows that $\langle\widetilde{S}_{22}\widetilde{S}_{22}\rangle =
\langle\widetilde{S}_{11}\widetilde{S}_{11}\rangle$ and
$\langle\widetilde{S}_{12}\widetilde{S}_{12}\rangle =
\frac{3}{4}\langle\widetilde{S}_{11}\widetilde{S}_{11}\rangle$.

Figure 2 shows isotropy ratios of the filtered gradient fields and of the
mean SGS stresses, for $\Delta/\eta=125$. In isotropic
conditions,  all ratios have to be on the unit line except 
$\langle\tau_{12}\rangle / \langle\tau_{11}\rangle$ which should tend to zero with
decreasing filter scale since the fraction of mean shear stress carried by the
SGS scales is expected to vanish when $\Delta/\ell \to 0$. As can be seen, both
the SGS stress and strain-rate fields are quite isotropic  across the heated
wake flow at this scale. For the larger filter sizes of $\Delta/\eta=250$ and
$500$ (not shown),  the trends are similar to figure 2 but slightly less isotropic.

Next, results are given for the SGS dissipation tensors across the wake.
We define the following `isotropy ratios':
\begin{equation}
I_{u22}  \equiv 
\frac{\left<\tau_{22}\widetilde{S}_{22}\right>}
{\left<\tau_{11}\widetilde{S}_{11}\right>},~~~
I_{u12}  \equiv 
\frac{\left<\tau_{12}\widetilde{S}_{12}\right>}
{\frac{3}{4}\left<\tau_{11}\widetilde{S}_{11}\right>},~~~
I_{\theta}  \equiv 
\frac{\left<q_{2}\widetilde{G}_{2}\right>}
{\left<q_{1}\widetilde{G}_{1}\right>}.
\end{equation}

\begin{figure}
\centerline{\hbox{
\psfig{figure=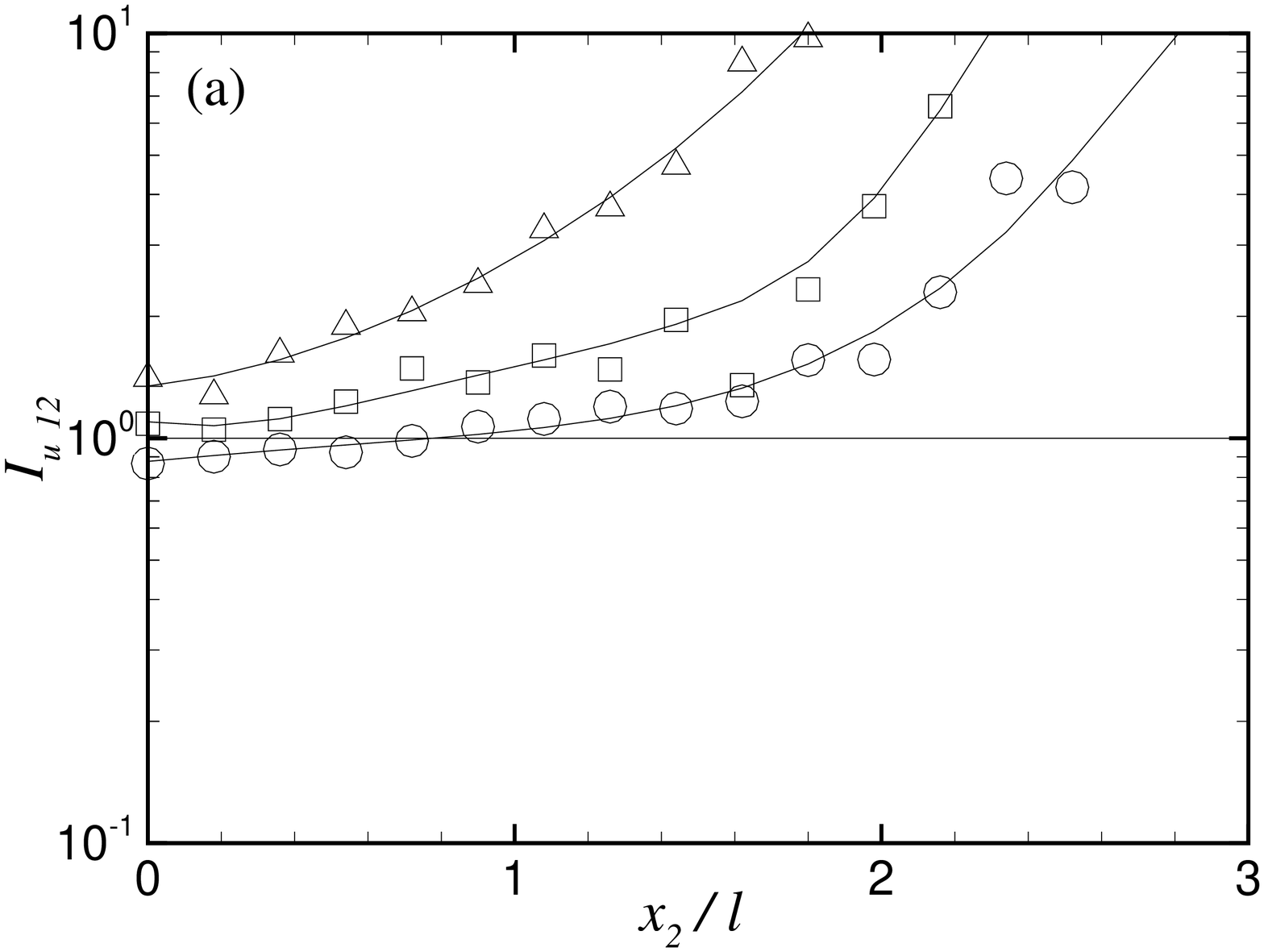,width=6.7cm,angle=0}
\psfig{figure=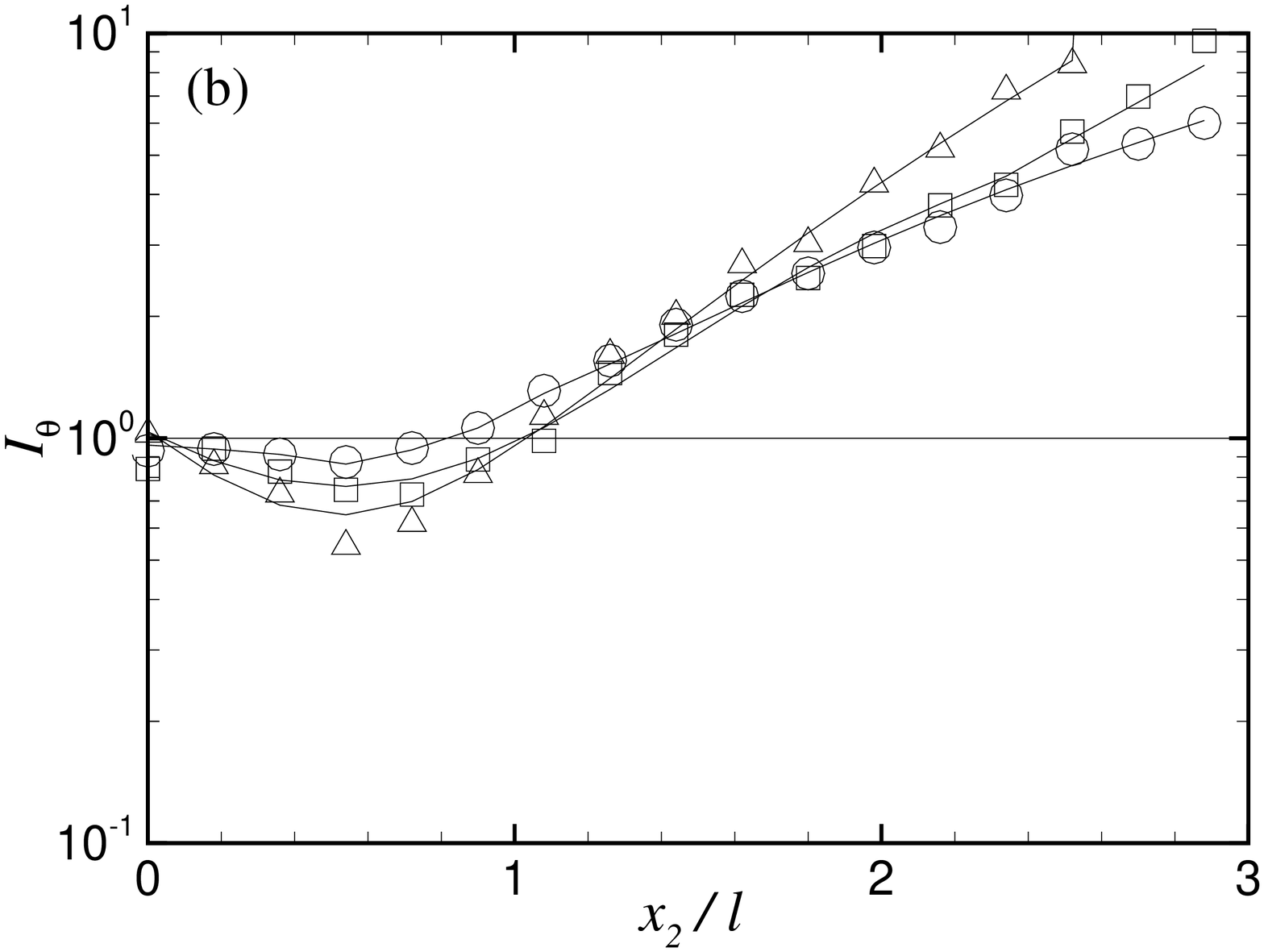,width=6.7cm,angle=0}}}
\caption{Isotropy ratios in the SGS dissipations:
(a) $I_{u12}$;
(b) $I_{\theta}$ (see Eq. 2).
The symbols represent:  
{\Large{$\circ$}}, $\Delta/\eta=125$;
$\square$, $\Delta/\eta=250$;
$\vartriangle$, $\Delta/\eta=500$. 
For clarity, $4^{th}$-order polynomial fits of $log I$ vs $x_2/\ell$ are
added as solid lines.
}
\label{<<Fig3>>}
\end{figure}

Figures 3(a) and (b) show the spatial distributions of the SGS kinetic energy dissipation 
isotropy ratios $I_{u12}$ and the SGS scalar-variance dissipation isotropy ratios 
$I_{\theta}$, respectively, across the wake flow with the different filter sizes. 
All moments in this study are statistically well converged. 
For example, running averages such as $n^{-1}\sum_{k=1}^n q_i \tilde{G}_j$ did not differ 
from their global average by more than 6\%
over the last 80\% of data (i.e. for $n\in[0.2N,N]$ where $N=2.4\cdot 10^6$).
For different filter sizes considered,
the SGS kinetic energy and scalar-variance dissipations are isotropic only
towards the centerline of the wake. Towards the edges of the wake, large
anisotropies persist (even though there the mean shear and mean temperature
gradients vanish also).  
The isotropy ratio of $I_{u12}$ at a fixed $x_2/\ell$ in figure 3(a) clearly increases 
with the filter size. 
However, the isotropy ratio of the scalar-variance dissipation 
$I_{\theta}$ shows almost no variation with filter size for $1 < x_2/\ell < 2$,
where the scalar gradient in the cross-wake direction, 
$\frac{\partial \theta}{\partial x_2}$, is large.

The increased anisotropy in the outer regions of the wake (see figures 3(a) and (b)) is 
another striking result. The anisotropy may be associated with the highly 
intermittent character of the flow there.
One may wonder whether the anisotropy arises from a superposition of distinct 
behaviors in the turbulent and the non-turbulent (outer) regions. Using conditional
averaging, O'Neil \& Meneveau (1997) already showed that the conditional averages of 
the SGS dissipations in the non-turbulent regions were negligible compared to 
those in the turbulent regions.  The implication was that the global averages of 
dissipation could be explained entirely by their conditional mean value inside 
the turbulent part: $\langle \tau_{ij}\tilde{S}_{ij}\rangle
\approx \Gamma \langle \tau_{ij}\tilde{S}_{ij}\rangle_T$, 
where $\Gamma(x_2)$ is the intermittency function (i.e. the fraction of time
the signal is turbulent at any given $x_2$), and the subscript ``$T$'' stands for 
averaging conditioned on `turbulence' 
(see O'Neil \& Meneveau 1997 for details).
When replacing the conditional averages multiplied by
$\Gamma$ in the expression for isotropy ratios, $\Gamma$ cancels from both
numerator and denominator. This behavior suggests that the anisotropy ratios
shown in figure 3 are equal to those inside the turbulent regions alone, and that
their rise in the outer parts of the wake cannot be explained by contributions
coming from the non-turbulent parts. 
However, it is still possible that 
non-trivial contributions to the scalar-variance dissipations could originate at 
the interface separating turbulent from non-turbulent regions.

\begin{figure}
\centerline{\hbox{
\psfig{figure=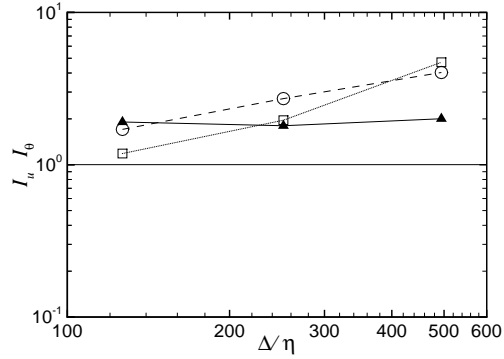,width=6.7cm,angle=0}}}
\caption{Isotropy ratios of kinetic energy and scalar-variance SGS dissipations 
at $x_2/\ell=1.44$, where the mean temperature gradient peaks.
The open symbols (and dashed lines) represent isotropy ratios for the SGS
dissipation of kinetic energy, namely {\Large{$\circ$}}, $I_{u22}$;
and $\square$, $I_{u12}$; the closed triangles $\blacktriangle$ (and solid line) are
for the scalar dissipation $I_{\theta}$.} 
\label{<<Fig4>>}
\end{figure}

To highlight the variations with filter scale more directly, in figure 4 we plot
the isotropy ratios as a function of scale, for the transverse location
$x_2/\ell=1.44$ close to the peak mean temperature gradient.
As is evident, the isotropy ratios of the kinetic energy dissipation 
decrease towards unity as the filter size decreases,
whereas the isotropy ratio of the scalar-variance dissipation remains 
unchanged near $I_{\theta} \sim 2$ as the filter size is decreased. 
As seen in figure 3, near the centerline, where the mean shear and 
scalar gradient vanish, the SGS isotropy ratios are all near unity. 

Therefore, we conclude that in terms of the most important of the resolved-SGS
scale interactions (the mean SGS dissipation), the scalar field maintains strong
anisotropy in the presence of a mean scalar gradient. Conversely, the
velocity field has the trends expected from approach to isotropy at small scales.

Next, we quantify the isotropy level of model predictions, when the
data are analyzed in an `a-priori' sense, i.e. by replacing
$\tau_{ij}$ and $q_j$ above by model expressions. First,  the  
standard eddy-diffusion model is considered, i.e., 
\begin{equation}
\tau_{ij}^{\mathrm{Smag}} - \frac{1} {3}\tau_{kk}^{\mathrm{Smag}}\delta_{ij} ~=~   
-2 (C_S\Delta)^2 |\widetilde{S}| \widetilde{S}_{ij}, ~~~~~
q_{j}^{\mathrm{Smag}}  ~=~  
-2 Pr^{-1}_{\mathrm{sgs}} (C_{S}\Delta)^2 |\widetilde{S}| \widetilde{G}_{j},
\end{equation}
where $|\widetilde{S}| = \big(2\widetilde{S}_{mn} \widetilde{S}_{mn} \big)
^{1/2}$ is the modulus of the resolved strain rate, $C_S$ is the Smagorinsky
coefficient, and $Pr_{\mathrm{sgs}}$ is the SGS Prandtl number.
Consequently, and independent of the model coefficients, the
isotropy ratios from the eddy-diffusion model are:
\begin{equation}
I_{u22}^{\mathrm{Smag}} \equiv 
\frac{\left<|\widetilde{S}|\widetilde{S}_{22} \widetilde{S}_{22}\right>}
{\left<|\widetilde{S}|\widetilde{S}_{11} \widetilde{S}_{11}\right>},~~~
I_{u12}^{\mathrm{Smag}} \equiv 
\frac{\left<|\widetilde{S}|\widetilde{S}_{12} \widetilde{S}_{12}\right>}
{\frac{3}{4}\left<|\widetilde{S}|\widetilde{S}_{11} \widetilde{S}_{11}\right>},~~~
I_{\theta}^{\mathrm{Smag}} \equiv 
\frac{\left<|\widetilde{S}|\widetilde{G}_{2} \widetilde{G}_{2}\right>}
{\left<|\widetilde{S}|\widetilde{G}_{1} \widetilde{G}_{1}\right>}.
\end{equation}
In computing the filtered strain-rate magnitude from the data, the
following approximation is used (this is a 2D extension of the 1D approach of
O'Neil \& Meneveau 1997, and was also used in Liu, Katz \& Meneveau 1999):
$
|\widetilde{S}| \approx 
\left[ 2 \left( 2\widetilde{S}_{11}\widetilde{S}_{11} +
\widetilde{S}_{22}\widetilde{S}_{22}+
6\widetilde{S}_{12}\widetilde{S}_{12} \right) \right]^\frac{1}{2}.
$
This approximation is not expected to affect the accuracy of our
measured isotropy ratios significantly. The reason is that $|\tilde{S}|$  and its 2D
approximation are scalars which multiply equally all terms of the squared velocity and
scalar gradients and is further supported by the isotropic behavior of second-order 
moments of $\tilde{S}_{ij}$ shown in figure 2.
The isotropy ratios of the SGS dissipations from the eddy-diffusion model are
shown in figure 5(a). Results are near
unity  almost independently of scale, including the passive scalar dissipation.
Profiles across the wake (not shown) also are near unity. This result is  
consistent with the observed isotropy of the  square gradient tensors shown in
figure 2 (the only difference here is the additional $|\widetilde{S}|$ factor). 
Hence, using the eddy-diffusion model one would (incorrectly) predict
SGS isotropy since the resolved second-order moments are isotropic.

\begin{figure}
\centerline{\hbox{
\psfig{figure=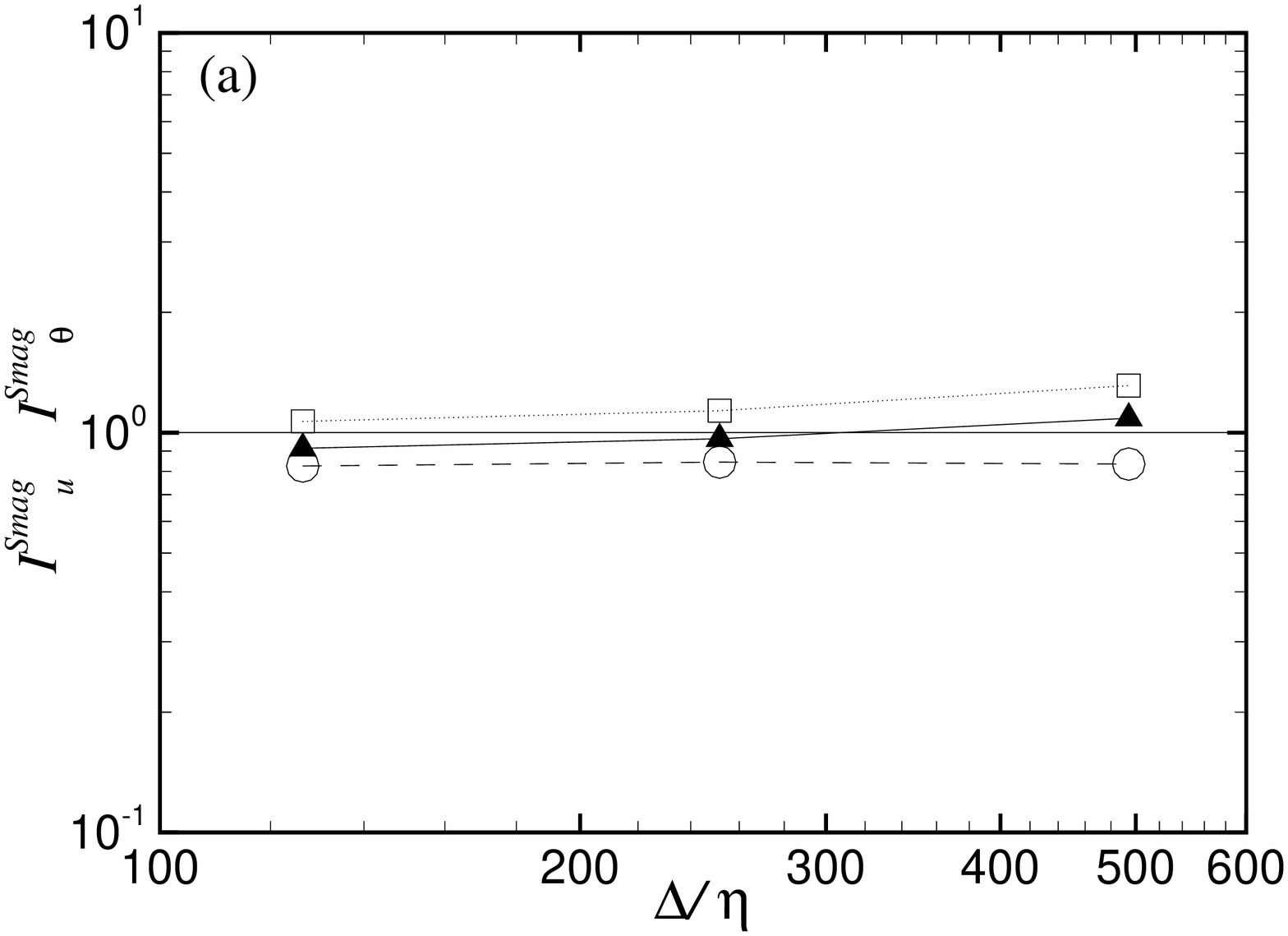,width=6.7cm,angle=0}
\psfig{figure=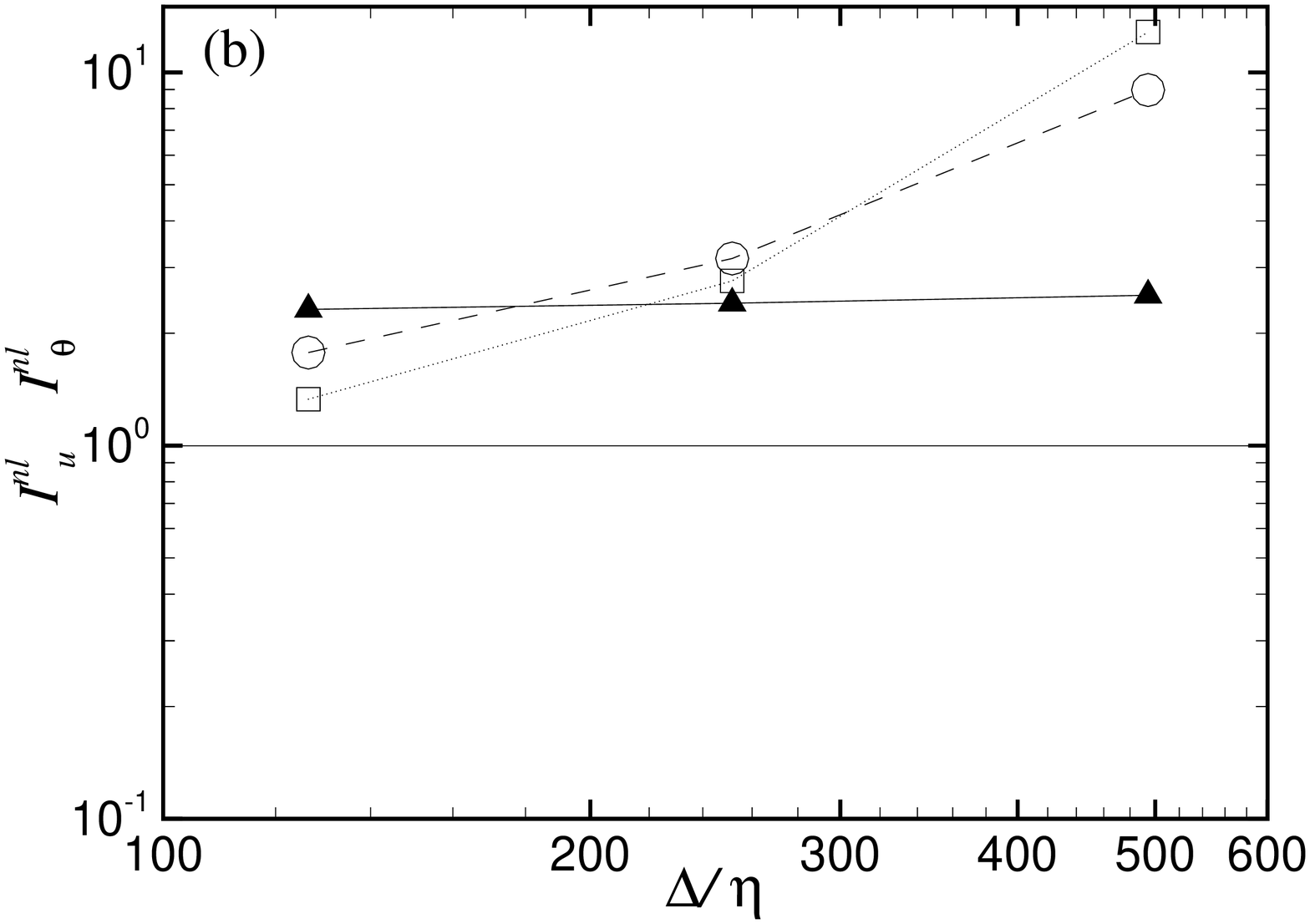,width=6.7cm,angle=0}}}
\caption{Isotropy ratios of components in SGS dissipations from models
at $x_2/\ell=1.44$, as a function of filter scale:
(a) eddy-diffusion models:
{\Large{$\circ$}}, $I_{u22}^{\mathrm{Smag}}$;
$\square$, $I_{u12}^{\mathrm{Smag}}$;
$\blacktriangle$, $I_{\theta}^{\mathrm{Smag}}$ (see Eq. 4);
(b) nonlinear models: 
{\Large{$\circ$}}, $I_{u22}^{\mathrm{nl}}$;
$\square$, $I_{u12}^{\mathrm{nl}}$;
$\blacktriangle$, $I_{\theta}^{\mathrm{nl}}$ (see Eq. 6).
}
\label{<<Fig5>>}
\end{figure}

The second modeling class to be considered here is the `nonlinear model' or 
`tensor eddy-diffusion model' (Leonard 1974,
Clark, Ferziger \& Reynolds 1979, Liu, Meneveau \& Katz 1994, 
Leonard 1997, Borue \& Orszag 1998
and Meneveau \& Katz 2000).  
The nonlinear model reads as follows:
\begin{equation}
\tau_{ij}^{\mathrm{nl}}  ~=~ 
C_{nl} \Delta^2 
\frac{\partial \widetilde{u}_i}{\partial x_k} 
\frac{\partial \widetilde{u}_j}{\partial x_k},~~~
q_{j}^{\mathrm{nl}}  ~=~ 
C_{{nl}_\theta} \Delta^2 
\frac{\partial \widetilde{\theta}}{\partial x_k} 
\frac{\partial \widetilde{u}_j}{\partial x_k},
\end{equation}
where
$C_{nl}$ and $C_{{nl}_\theta}$ are the nonlinear model coefficients. 
Therefore, the corresponding isotropy ratios from the nonlinear model are defined
as follows:
\begin{equation}
I_{u22}^{\mathrm{nl}} \equiv 
\frac{\left<\frac{\partial \widetilde{u}_2}{\partial x_k}
\frac{\partial \widetilde{u}_2}{\partial x_k}\widetilde{S}_{22}\right>}
{\left<\frac{\partial \widetilde{u}_1}{\partial x_k}
\frac{\partial \widetilde{u}_1}{\partial x_k}\widetilde{S}_{11}\right>},~~~
I_{u12}^{\mathrm{nl}} \equiv 
\frac{\left<\frac{\partial \widetilde{u}_1}{\partial x_k}
\frac{\partial \widetilde{u}_2}{\partial x_k}\widetilde{S}_{12}\right>}
{\frac{3}{4} \left<\frac{\partial \widetilde{u}_1}{\partial x_k}
\frac{\partial \widetilde{u}_1}{\partial x_k}\widetilde{S}_{11}\right>},~~~
I_{\theta}^{\mathrm{nl}} \equiv 
\frac{\left<\frac{\partial \widetilde{\theta}}{\partial x_k}
\frac{\partial \widetilde{u}_2}{\partial x_k}\widetilde{G}_{2}\right>}
{\left<\frac{\partial \widetilde{\theta}}{\partial x_k}
\frac{\partial \widetilde{u}_1}{\partial x_k}\widetilde{G}_{1}\right>},
\end{equation}
where $k$ varies from 1 to 2 in the analysis of our 2D data.

The isotropy ratios of the SGS dissipations from the nonlinear model are
shown in figure 5(b), as a function of filter scale. As is apparent in comparing with
figure 4, the main features of anisotropy and filter-size dependence are
correctly reproduced qualitatively. Quantitatively, the levels of anisotropy are
overestimated. For instance, the level of anisotropy for the modeled
SGS scalar-variance dissipation appears to stay near $I_{\theta}^{\mathrm{nl}}  
\sim 2.5$ as opposed to $I_{\theta} \sim 2$ for the real SGS scalar dissipation.

\section{Summary and conclusions}
In studying passive scalar statistics in a turbulent shear flow with a mean
temperature gradient we focus on statistics of interest to subgrid
modeling and large-eddy simulation. In order to obtain the filtered and subgrid
velocities and temperatures, a probe array composed of four X-wire and four
cold-wire sensors is used and two-dimensional box filtering in the streamwise and
cross-wake directions is applied to the data. The isotropy ratios of the
SGS kinetic energy and scalar-variance dissipations are
investigated as function of position in the flow and as a function of filter scale.
Both dissipations are isotropic independent of filter
size near the centerline  where there is no mean shear or scalar gradient.
However, at locations with high gradient (and also further towards the outer wake
regions), we find that the scalar-variance dissipation remains
highly anisotropic, independent of filter size. Conversely, the kinetic energy
dissipation tensor approaches isotropy as the filter size is decreased.
A mechanistic explanation of the observed trends in terms of possible
orientations of ramp and cliff structures is not evident to us at this time.
The persistence of scalar anisotropy even at small scales is consistent
with prior results for structure functions and gradient statistics of 
unfiltered turbulence (Mestayer 1982, Sreenivasan 1991, Warhaft 2000). 
Present results quantify the impact of  this anisotropy on the interactions among 
large and small scales in the context of SGS modeling and LES.

We find that the predictions of eddy-diffusion models are much more isotropic than
the real phenomenon. This may at first glance seem obvious since eddy-diffusion
models are often invoked under the banner of isotropy. However, the
present result does not arise from an explicit isotropy assumption built into the model, 
but because the resolved gradients have second-order statistics that are isotropic. 
For instance, if the scalar gradient tensor $\langle\tilde{G}_i\tilde{G}_j\rangle$  
had been found to be anisotropic, it would have implied anisotropic behavior of the
eddy-diffusion model's predictions.
Instead, the isotropy that exists in the filtered scalar gradients is
incorrectly applied to model the anisotropic statistics of the SGS heat flux.  
The main problem for the eddy-diffusion model appears to be that it uses second-order 
statistics to model third-order statistics. Conversely, the anisotropy exists 
in the third-order moments  that arise from the velocity
and scalar product ($q_i$) multiplied by the scalar gradient ($\widetilde{G}_j$), but
is not discernible in the second-order statistics of
$\widetilde{G}_j$ alone (even when modulated by the strain-rate magnitude). 
The anisotropy is clearly discernible, however, in the third-order moments consisting of
the filtered velocity gradients times scalar gradients squared that arise in the
expression for modeled SGS dissipation of scalar-variance using the nonlinear model.
These expressions are able to reproduce the detailed phase relationships among the
velocity and scalar field that govern the SGS dissipation (cascade) of scalar
variance. We remark that the overprediction of anisotropy by the nonlinear model (and
its underprediction by the eddy-diffusion model) is  reminiscent of the opposing
trends of these two models documented in Liu \etal (1999) for rapidly strained 
turbulence in cold-flow.  The opposing trends suggest that a linear combination of
the two models, i.e. the `mixed model', can be tuned to reproduce the correct
amount of anisotropy (for a discussion of the application of mixed models in LES,
see Meneveau \& Katz 2000).  However, note that the data for $I_{u12}$ 
and $I_{u22}$ at $\Delta/\eta=250$ and 125 do not provide clear justification 
for choosing the mixed over the non-linear model.

Finally, it is stressed that current results are obtained in a single flow for a 
single moderate Reynolds number.  Even if the present evidence in figure 4
of an essentially scale-independent anisotropy for the scalar dissipation appears 
to be quite strong, the results could change in another flow, 
or at higher Reynolds numbers (we recall that according to Sreenivasan 1996, universal 
behavior for the scalar requires $Re_{\lambda}$ above 1000 or so).  These considerations 
serve as motivation for further work in this area.

\acknowledgements
We thank Professors Z. Warhaft and L. Mydlarski for useful comments and
suggestions about the cold-wire calibration. This work was financially supported by
the National Science Foundation (grant CTS-9803385).


\end{document}